**Inversion of DC Resistivity Data using Physics-Informed Neural Networks**

Rohan Sharma[*1], Divakar Vashisth[2], Kuldeep Sarkar[1] and Upendra Kumar Singh[1]

[1]*Department of Applied Geophysics, Indian Institute of Technology Dhanbad*, [2]*Department of Energy Science and Engineering, Stanford University*

email: * rohanlatha29@gmail.com

## Summary

The inversion of DC resistivity data is a widely employed method for near-surface characterization. Recently, deep learning-based inversion techniques have garnered significant attention due to their capability to elucidate intricate non-linear relationships between geophysical data and model parameters. Nevertheless, these methods face challenges such as limited training data availability and the generation of geologically inconsistent solutions. These concerns can be mitigated through the integration of a physics-informed approach. Moreover, the quantification of prediction uncertainty is crucial yet often overlooked in deep learning-based inversion methodologies. In this study, we utilized Convolutional Neural Networks (CNNs) based Physics-Informed Neural Networks (PINNs) to invert both synthetic and field Schlumberger sounding data while also estimating prediction uncertainty via Monte Carlo dropout. For both synthetic and field case studies, the median profile estimated by PINNs is comparable to the results from existing literature, while also providing uncertainty estimates. Therefore, PINNs demonstrate significant potential for broader applications in near-surface characterization.

## Introduction

The direct-current (DC) resistivity method is a widely utilized technique in near-surface geophysics, attributed to its high sensitivity to variations in the resistivity of geological features and its cost-effectiveness. It has been popularly employed in groundwater investigation (Benson et al., 1997) and mineral exploration (Srigutomo et al., 2016). In DC resistivity surveys, an electric current is injected into the ground, and the potential difference is measured to calculate the apparent resistivity using Ohm's law. To obtain the true resistivity model of the subsurface and facilitate interpretation, this apparent resistivity data is inverted. Some researchers have tried solving this inverse problem using global optimization methods like simulated annealing (Sen et al., 1993), genetic algorithm (Liu et al., 2012), and variable weight particle swarm optimizer - grey wolf optimizer (Sarkar et al., 2023).

The advent of machine learning algorithms has proven effective in geophysical inversion, enabling complex nonlinear mappings between observed geophysical data and subsurface earth model parameters. In a supervised setting, they have been employed to invert DC resistivity data (El-Qady and Ushijima, 2001; Singh et al., 2019; Aleardi et al., 2021). However, inversion based on supervised deep learning presents challenges, such as the limited availability of training data and estimations that could be geologically and physically inconsistent. Additionally, there is a need for uncertainty quantification in the predicted model parameters. These considerations are driving the adoption of physics-informed methodologies in the field of geophysical inversion (Vashisth and Mukerji, 2022; Liu et al., 2023a; Liu et al., 2023b).

This paper introduces a convolutional neural network (CNN) based, physics-informed deep learning framework tailored for unsupervised 1D DC resistivity data inversion, drawing inspiration from Vashisth and Mukerji (2022). Our primary contribution is the development of a lightweight CNN architecture, applied to both case studies and employed to estimate uncertainties in the predicted resistivities through Monte Carlo dropout. In the following sections, we outline our employed methodology and demonstrate the efficacy of our approach through both synthetic and field case studies.

**Methodology**

The DC resistivity Physics Informed Neural Network (PINN) architecture comprises of an encoder and a decoder (Figure 1). The encoder consists of three convolutional blocks, each containing a 1D convolutional layer, a rectified linear units (ReLU) layer for non-linearity, and a dropout layer with a dropout rate of 0.2 for regularization and uncertainty estimation. The outputs from the final convolutional block are flattened and fed into a fully connected layer with the number of output nodes equal to the number of layered earth model parameters. To define an n-layered earth model, we require 2n-1 model parameters, consisting of the resistivity ($\rho$) and thickness (h) of each layer, with the bottommost layer assumed to be semi-infinite. The fully connected layer's output is passed through a sigmoid activation function to constrain the values between 0 and 1. Subsequently, these values are rescaled to fall within the specified bounds (search space) for resistivities and thicknesses before being fed into the decoder.

The output of the encoder serves as the input for the decoder, which incorporates the physics of Schlumberger sounding (Koefoed, 1979). The decoder generates apparent resistivity curves corresponding to the input resistivity profiles. The training phase makes use of the Adam optimizer with an initial learning rate of 0.001, aiming to minimize the root mean squared (RMS) error between the logarithms of the observed and predicted apparent resistivity curves. It is important to note that incorporating the logarithm of the apparent resistivity values into the loss function was crucial for achieving better fits, particularly in cases with highly varying resistivities. The learning rate was reduced by a factor of 0.8 if there was no improvement in the loss for 15 consecutive epochs. Upon completion of the training process, the encoder's output vector acquires meaningful representations in the form of the learned layered earth model parameters. This allows the encoder to directly predict resistivity profiles from the apparent resistivity curves. The values of all hyperparameters within the architecture were determined through an extensive exploration of various combinations.

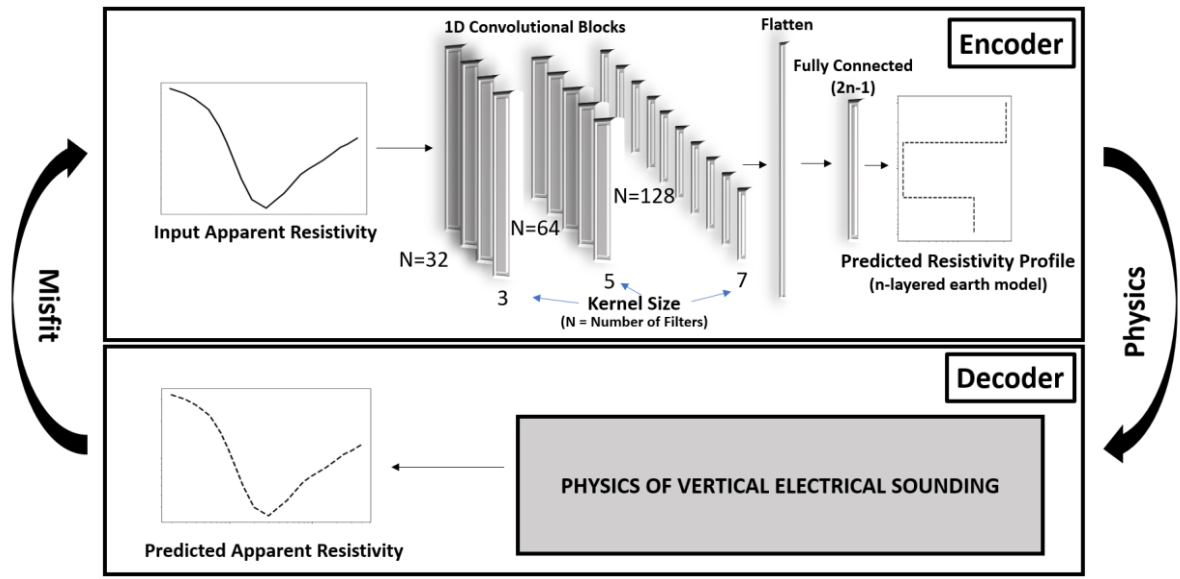

*Figure 1* Unsupervised DC Resistivity PINN architecture. The encoder is a CNN trained to generate resistivity profile from a given input apparent resistivity curve. The decoder incorporates the physics of Schlumberger sounding to guide the encoder's learning process.

During the prediction phase, dropout is still employed to mimic the behaviour observed during training. This approach, known as Monte Carlo Dropout, serves to approximate the posterior distribution of the model parameters, thereby aiding in uncertainty estimation (Gal and Ghahramani, 2016). By repeating this process multiple times (in this case, 10000 times), empirical samples of the predictions are generated. These samples facilitate the computation of approximate posterior percentiles. Notably, the 50th percentile (P-50) represents the median value, while the range between the 5th (P-5) and 95th percentile (P-95) provides an estimate of the network's prediction uncertainty (Das and Mukerji, 2019).

## Results and Discussion

In this study, we utilize the same vertical electric sounding (VES) synthetic and field dataset as presented in Sarkar et al. (2023). Our analysis involves comparison of the results obtained through

PINNs with those derived from the application of the variable weight particle swarm optimizer - grey wolf optimizer (vPSOGWO) algorithm, as demonstrated in Sarkar et al. (2023).

The synthetic apparent resistivity dataset, incorporating 10% Gaussian noise (Figure 2a), is derived from a three-layered earth model (Figure 2b) with resistivities conforming to an H-type curve. Figure 2b demonstrates that the median (P-50) of PINNs predicted resistivity profiles closely matches the true profile (Figure 2b), with the computed apparent resistivity curve for this profile (Figure 2a) exhibiting an RMS error of 16.55 $\Omega$m, significantly outperforming the 45.39 $\Omega$m RMS error associated with vPSOGWO algorithm. Furthermore, the uncertainty in the predicted resistivity profiles (spread between P-5 and P-95) remains relatively constant with depth, highlighting consistency in the estimations.

For the field case study (Figure 3), VES data from Digha, Medinipur (West Bengal) is inverted. The geological profile for the area is characterized by alternating layers of sedimentary rock containing sand and clay, overlaid by a thin alluvial layer. The apparent resistivity curve (Figure 3a) corresponding to the PINNs estimated P-50 resistivity profile (Figure 3b) exhibits a strong alignment with the observed data, yielding an RMS error of 0.49 $\Omega$m, which is comparable to the 0.50 $\Omega$m RMS error given by vPSOGWO algorithm. Notably, the uncertainty in the PINNs predicted resistivity profiles decreased considerably below the estimated conductive alluvial layer, indicating a higher level of confidence in the model's predictions in this region.

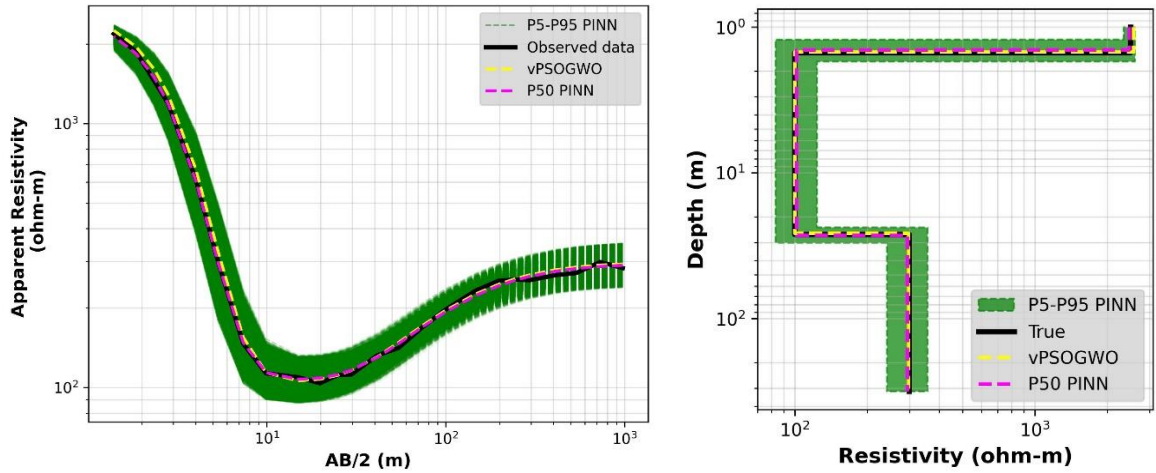

***Figure 2*** *(a) The observed (black) and computed DC apparent resistivity curves for the synthetic case study. (b) The true (black) and estimated resistivity profiles from PINNs and vPSOGWO. The PINNs estimated median resistivity profile (P-50) and its corresponding apparent resistivities are illustrated in pink, while the predictions highlighting the uncertainty range (between P-5 and P-95) are shown in green. The vPSOGWO estimated resistivity profile and apparent resistivities are shown in yellow.*

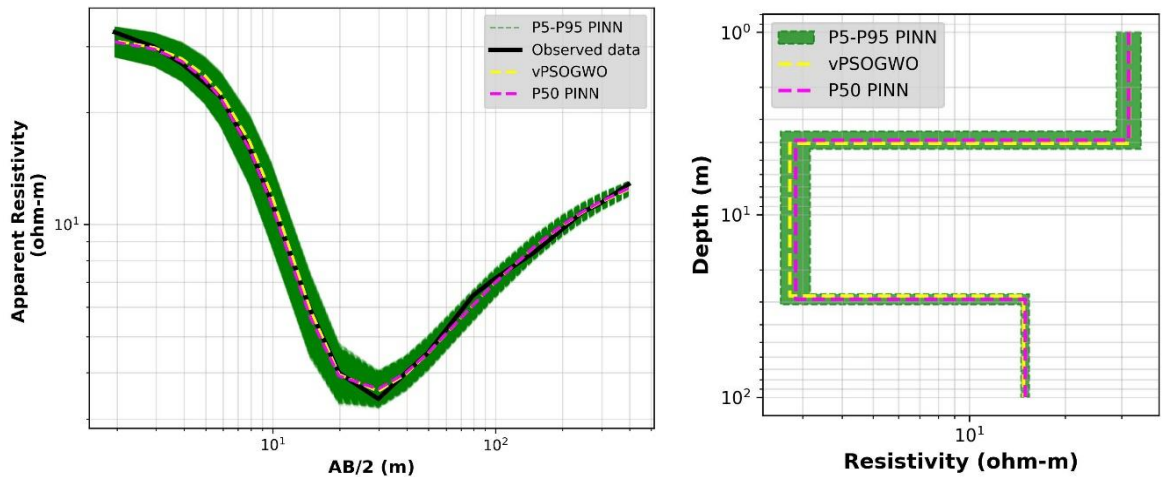

***Figure 3*** *(a) The observed (black) and computed DC apparent resistivity curves for the Digha field case study. (b) The estimated resistivity profiles from PINNs and vPSOGWO. The PINNs estimated median resistivity profile (P-50) and its corresponding apparent resistivities are illustrated in pink, while the predictions highlighting the uncertainty range (between P-5 and P-95) are shown in green. The vPSOGWO estimated resistivity profile and apparent resistivities are shown in yellow.*

The findings from both synthetic and field case studies underscore the robustness and accuracy of the proposed PINNs architecture for inverting DC resistivity data, with all observed apparent resistivity data points falling within the model-predicted uncertainty bounds. Future work will explore the application of PINNs to the inversion of 2D and 3D Electrical Resistivity Tomography (ERT) data.

## Conclusions

This study highlights the efficacy of the proposed PINNs architecture in accurately inverting DC resistivity data, showcasing performance comparable to that reported in existing literature. The architecture proved to be robust in handling both synthetic and field datasets, predicting consistent resistivity profiles and providing reliable uncertainty estimates. These results underscore the potential of PINNs as a reliable and valuable tool for geophysical data inversion tasks and near-surface characterization.